\documentclass[11pt]{article}

\usepackage{geometry}  % Flexible and complete interface to document dimensions
\usepackage[T1]{fontenc}  % Standard package for selecting font encodings
\usepackage[utf8x]{inputenc}  % Accept different input encodings
\usepackage{enumitem}  % Control layout of itemize, enumerate, description
\usepackage{titling}  % Control over the typesetting of the \maketitle command
\usepackage[numbers,super,comma,sort&compress]{natbib}  % Flexible bibliography support
\usepackage{etoolbox}
\usepackage{mathtools}  % Mathematical tools to use with amsmath
\usepackage{titlesec}  % Select alternative section titles
\usepackage{gensymb}   % Degree symbol
\usepackage[english]{babel}  % Multilingual support for Plain TeX or LaTeX
\usepackage{amsmath}  % AMS mathematical facilities for LaTeX
\usepackage{amsfonts}  % TeX fonts from the American Mathematical Society
\usepackage{amssymb}  % Additional symbols from American Mathematical Society
\usepackage{wasysym}  % LaTeX support file to use the WASY2 fonts
\usepackage{bbm}  % "Blackboard-style" cm fonts
\usepackage{array}  % Extending the array and tabular environments
\usepackage{xr}  % References to other LaTeX documents
\usepackage{verbatim} % Reimplementation of and extensions to LaTeX verbatim
\usepackage{float} % Improved interface for floating objects
\usepackage{caption}
\usepackage{subcaption}

\titleformat{name=\section}{\normalfont\Large\bfseries}{}{0pt}{}
\titleformat{name=\subsection}{\normalfont\large\bfseries}{}{0pt}{}
\titleformat{name=\subsubsection}{\normalfont\normalsize\bfseries}{}{0pt}{}

\newcommand{\optincludegraphics}[2][]{}
\newcommand{\optinput}[1]{}
\newtagform{brackets}{[}{]}
\usetagform{brackets}

\title{Estimating air quality co-benefits of energy transition using machine learning}

\immediate\write18{texcount -sub=none -merge -incbib -dir -utf8 main.tex | grep _Abstract_ | grep -oE '[0-9]+' | python -c "import sys; print(sum(int(l) * w for l, w in zip(sys.stdin, [1, 1, 0, 0, 0, 0, 0])))" > \jobname.wcAbstract }

\immediate\write18{texcount -sub=none -merge -incbib -dir -utf8 main.tex | grep _main_ | grep -oE '[0-9]+' | python -c "import sys; print(sum(int(l) * w for l, w in zip(sys.stdin, [1, 1, 0, 0, 0, 0, 0])))" > \jobname.wcManuscript }

\immediate\write18{texcount -sub=none -merge -incbib -dir -utf8 main.tex | grep _Methods_ | grep -oE '[0-9]+' | python -c "import sys; print(sum(int(l) * w for l, w in zip(sys.stdin, [1, 1, 0, 0, 0, 0, 0])))" > \jobname.wcMethods }

\newcommand{\wcAbstract}{\input{\jobname.wcAbstract}}
\newcommand{\wcManuscript}{\input{\jobname.wcManuscript}}
\newcommand{\wcMethods}{\input{\jobname.wcMethods}}

\begin{document}
% ======================================================================

\begin{titlepage}
{\noindent\LARGE\bf\thetitle}

\bigskip

% : Insert author names, affiliations and corresponding author email
%FIXME
\begin{flushleft}\large
	Da Zhang\textsuperscript{1,2},
	Qingyi Wang\textsuperscript{3},
	Shaojie Song\textsuperscript{4},
	Simiao Chen\textsuperscript{5,6},
    Mingwei Li\textsuperscript{7},
    Lu Shen\textsuperscript{4},
	Siqi Zheng\textsuperscript{8},
    Bofeng Cai\textsuperscript{9,{*}},
    Shenhao Wang\textsuperscript{8,10,{*}}
\end{flushleft}

\bigskip

\noindent
%FIXME
\begin{enumerate}[label=\textbf{\arabic*}]
\item Institute of Energy, Economy, and Environment, Tsinghua University, Beijing, China
\item Joint Program on the Science and Policy of Global Change, Massachusetts Institute of Technology, Cambridge, MA, USA
\item Department of Civil and Environmental Engineering, Massachusetts Institute of Technology, Cambridge, MA, USA
\item School of Engineering and Applied Sciences, Harvard University, Cambridge, MA, USA
\item Heidelberg Institute of Global Health, Faculty of Medicine and University Hospital, Heidelberg University, Heidelberg, Germany
\item Chinese Academy of Medical Sciences and Peking Union Medical College, Beijing, China
\item Center for Policy Research on Energy and the Environment, Princeton University, Princeton, NJ, USA
\item Department of Urban Studies and Planning, Massachusetts Institute of Technology, Cambridge, MA, USA
\item Center for Climate Change and Environmental Policy, Chinese Academy for Environmental Planning, Beijing, China
\item Media Lab, Massachusetts Institute of Technology, Cambridge, MA, USA, 02139

%\item Department of Earth, Atmospheric, and Planetary Sciences, Massachusetts Institute of Technology, Cambridge, MA, USA, 02139
%\item Institute for Data, Systems, and Society, Massachusetts Institute of Technology, Cambridge, MA, USA, 02139
%\item Sloan School of Management, Massachusetts Institute of Technology, Cambridge, MA, USA, 02139
\end{enumerate}

\bigskip

% : Use the dagger symbol to denote a single equal contribution authorship.
% : Multiple equal-contribution authorship may be included in the acknowledgments.
%FIXME
%\textbf{{†}}: These authors contributed equally to this work.

% : Use the asterisk to denote corresponding authorship.
%FIXME
\textbf{*} To whom correspondence should be addressed. E-mail: shenhao@mit.edu and caibf@caep.org.cn.
\vfill

% ======================================================================
% : set word count results
% : to use `texcount` results, use '%TC:ignore'/'%TC:endignore' directives.
% : Exclude this part in submission
% \wordcount{\wcAbstract}{\wcManuscript}{\wcMethods}

\end{titlepage}

% ======================================================================
\pagebreak
% ======================================================================

% : start line numbers from here
% \linenumbers

%TC:break _Abstract_
% ======================================================================
\section{Abstract}
% ======================================================================

Estimating health benefits of reducing fossil fuel use from improved air quality provides important rationales for carbon emissions abatement. Simulating pollution concentration is a crucial step of the estimation, but traditional approaches often rely on complicated chemical transport models that require extensive expertise and computational resources. In this study, we develop a novel and succinct machine learning framework that is able to provide precise and robust annual average fine particle (PM$_{2.5}$) concentration estimations directly from a high-resolution fossil energy use data set. The accessibility and applicability of this framework show great potentials of machine learning approaches for integrated assessment studies. Applications of the framework with Chinese data reveal highly heterogeneous health benefits of reducing fossil fuel use in different sectors and regions in China with a mean of \$34/tCO$_2$ and a standard deviation of \$84/tCO$_2$. Reducing rural and residential coal use offers the highest co-benefits with a mean of \$360/tCO$_2$. Our findings prompt careful policy designs to maximize cost-effectiveness in the transition towards a carbon-neutral energy system.

%TC:break _main_
% ======================================================================
% ======================================================================
%\section{}
% ======================================================================

\newpage
The use of fossil fuel is well known as the major source of both greenhouse gas (GHG) and air pollutant emissions in most nations. Emissions abatement from fossil-fuel use reduction therefore could bring significant health benefits while addressing climate change{\cite{McCollum2013,West2013,Thompson2014,Driscoll2015,Shindell2016,Li2018,Shindell2018,Vandyck2018,Mayfield2019}}. Understanding the heterogeneity of local health benefits from marginal emissions reductions is crucial for designing co-control abatement activities. To obtain the geographical distribution of benefits with high resolution, it usually requires adjoint{\cite{Dedoussi2020}}, or more commonly, reduced-form chemical transport models (CTMs) that can be iterated many times to simulate air quality changes{\cite{MullerMendelsohn2009,Holland2016,Tessum2019,Goodkind2019,Hill2019}}. These reduced-form air quality models, e.g., the Air Pollution Emission Experiments and Policy analysis model (APEEP) {\cite{Muller2011}} and its updated versions (AP2/AP3), the Estimating Air quality Social Impacts Using Regression (EASIUR){\cite{Heo2016}}, the Intervention Model for Air Pollution (InMAP){\cite{Tessum2017}}, and the TM5-Fast Scenario Screening Tool (TM5-FASST){\cite{VanDingenen2018}}, often rely on linearized representations of emissions-concentration sensitivity derived from full-scale CTMs. 

Although reduced-form models provide satisfying approximations for full-scale models, several important limitations restrict their use in policy analysis. First, inputs for the reduced-form model that are developed from the full-scale model (usually for a specific simulation period) are still resource-intensive, hence updating the meteorological conditions that dictate the simulation results requires substantial domain knowledge and efforts. Second, for regions where the atmospheric chemistry processes are less well understood, especially the developing nations, the biases embodied in the full-scale model will be inherited by the reduced-form model, hence affecting the accuracy of policy simulations. Third, similar to the full-scale model, the reduced-form model also needs a detailed list of emissions inventory for all the pollutant species as inputs, so strong assumptions may have to be imposed to convert fossil fuel use data to emissions inventory. To address these limitations, in this paper, we develop a machine-learning framework that is able to provide precise and robust annual average fine particle (PM$_{2.5}$) concentration estimations directly from high-resolution fossil energy use data and some additional geographic information. We apply this framework with data from China and estimate health benefits of reducing fossil fuel use in different sectors and regions. The framework is transparent, easy-to-update, and ready to be extended to other nations without establishing sophisticated full-scale CTMs.

Our framework also supplements the growing literature that applies machine learning approaches to study air-quality-related topics. Several studies have exploited the strong data-fitting capacity of machine learning methods to predict short-term pollutant concentrations. For example, Ong et al.{\cite{Ong2016}} use a deep recurrent neural network (DRNN) to predict next 12-hour PM$_{2.5}$ concentrations in 52 Japanese cities; Kerckhoffs et al.{\cite{Kerckhoffs2019}} compare performances of different machine learning methods (e.g., bagging and random forest) to predict average of three 24-h measurements of ultrafine particles (UFP) using mobile and short-term stationary measurements in Dutch cities; Xing et al.{\cite{Xing2020}} develop a deep-learning-based response surface model (DeepRSM) that is trained using the Community Multiscale Air Quality (CMAQ) simulations on domains that cover China to characterize the response of O$_3$ and PM$_{2.5}$ concentrations to emissions changes; Kelp et al.{\cite{Kelp2020}} discuss the design of stable, general machine-learned models of the atmospheric chemical system. There are also studies{\cite{Li2017AE,Li2017GRL}} that utilize machine learning methods, for example, deep belief network (DBN) and generalized regression neural network (GRNN), to predict long-term (seasonal or annual) PM$_{2.5}$ concentrations using aerosol optical depth (AOD) data. Despite these existing studies, our paper applies machine learning approaches to estimate long-term air quality directly from emissions sources data (i.e., energy use) as well as health co-benefits of fossil energy use and CO$_2$ emissions for a large country for the first time, providing highly relevant implications for policy making. 

% ======================================================================
\section{Results}
% ======================================================================

\textbf{Structure and performance of the machine-learning framework} 

Our novel machine-learning framework applies a modified convolutional neural network architecture (ResCNN) to simulate annual average fine particle (PM$_{2.5}$) concentration observed by all China's national-level air-quality monitoring stations in 2015 (1,497 stations in total and 943 are used in the model due to some monitors overlapping in the same grid cell at the geographic resolution we use -- 10$km$ $\times$ 10$km$) from high-resolution fossil energy use data and some additional geographic information. This resolution (100 $km^{2}$) is finer than most regional reduced-formed models, for example, 1,642 $km^{2}$ (median) for AP2{\cite{Muller2011}}, 1,296 $km^{2}$ (median) for EASIUR{\cite{Heo2016}}, 293 $km^{2}$ (population-weighted mean) for InMAP{\cite{Tessum2017}}, and comparable to some regional comprehensive CTMs, for example, 144 $km^{2}$ for many CTMs for the United States{\cite{Goodkind2019}}. We use the most recent year with available data (year 2015) for the analysis in this paper, but the framework could be easily extended to future years and other regions. Specifically, the model output is annual average PM$_{2.5}$ concentration for each monitoring station collected from China's National Urban Air Quality Real-time Disclosure Portal{\cite{AQRealtime}}. The model inputs for predicting the annual average PM$_{2.5}$ concentration of a specific monitoring station are 2D tensors that contain geographical distribution of energy use (by sector and energy type) as well as altitude, temperature, and precipitation in a large surrounding area centered by the station (610 $km$ $\times$ 610 $km$, comparable to the domain size used for regional air quality modeling{\cite{Zhang2015}}). 

\begin{figure*}[ht!]
\centerline{\includegraphics[width=0.5\textwidth]{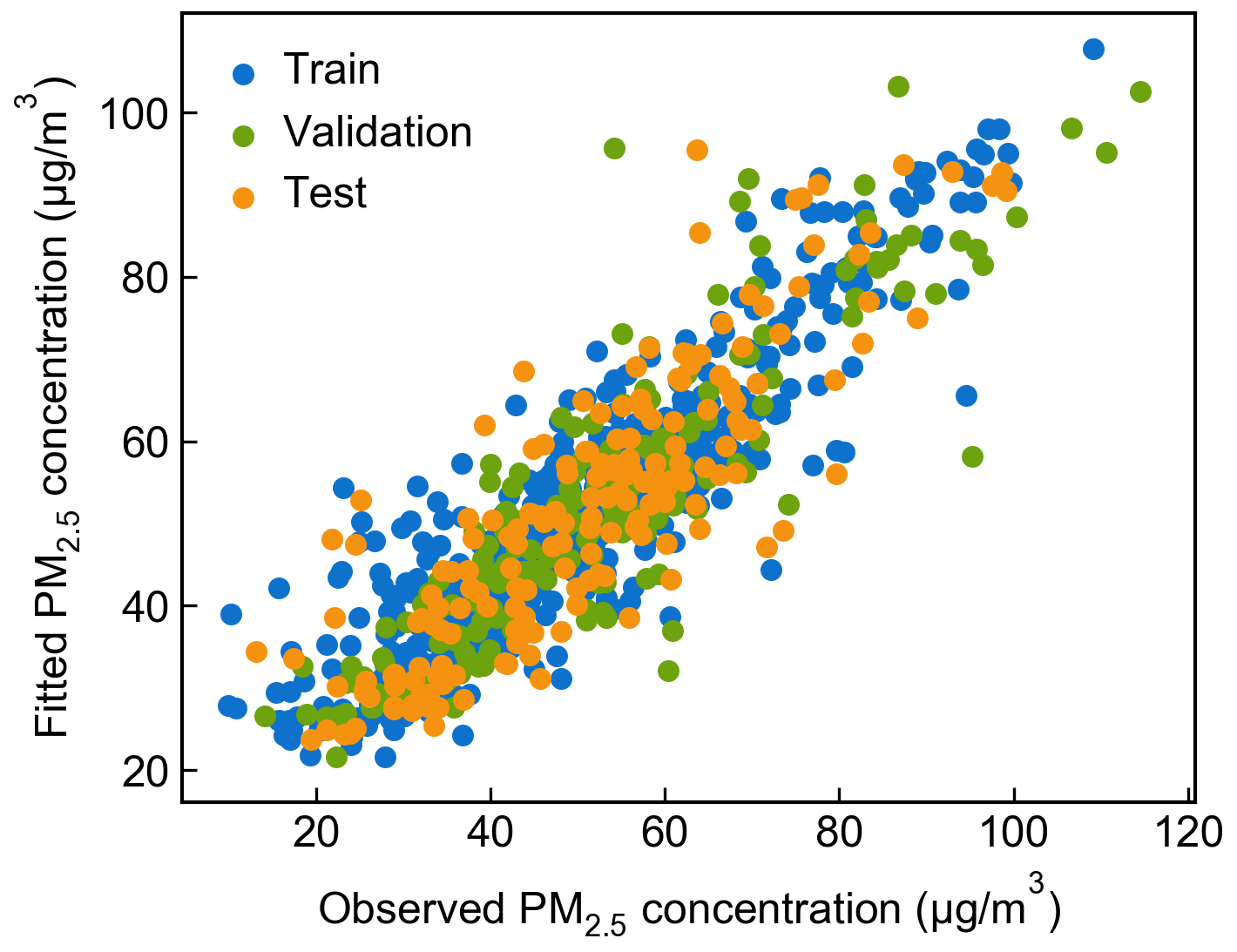}}
\caption{Comparison of model-predicted and observed PM$_{2.5}$ concentrations at 943 stations in mainland China for year 2015 by training, validation, and test data sets.}
\label{fig:fig1}
\end{figure*}

By adopting a standard training process, we select optimal hyper-parameters and correspondingly optimized parameters of the ResCNN framework. We show the prediction performance of our model by comparing the observed and model-predicted concentrations at 943 stations in mainland China for year 2015 (see \textbf{Figure \ref{fig:fig1}}). The training set shows the best fit but the validation and test set are also reasonably fitted, suggesting over-fitting is not a major issue. For the test set, mean fractional bias (MFB), mean fractional error (MFE), mean proportional error (MPE), correlation coefficient ($\rho$), and R square ($R^2$) are -0.04, 0.12, 0.06, 0.88, and 0.75, respectively, significantly improved compared to the fit of some published reduced-form air quality models (MFB: $-0.06$, MFE: 0.36, $\rho$: 0.74, $R^2$: 0.13 in Goodkind et al.{\cite{Goodkind2019}}; MPE: 0.37, $\rho$: 0.62 in Muller{\cite{Muller2011}}, see details in \textbf{Supplementary Table 2}). The strong predictive power of our model offers us a considerable advantage over conducting the ensuing air quality co-benefits estimation.

We then simulate annual average PM$_{2.5}$ concentration changes under different policy scenarios with reduced fossil energy use. In this study, we focus on the industrial coal use (including coal use in the power sector), road transportation oil use, and rural and residential coal use (mainly for heating and cooking), as they are the main targets of energy and environmental policies in China{\cite{Zhang2019}}. For illustration purposes, we first simulate scenarios that curtail fossil energy use in the above sectors by a certain percentage (from 2\% to 20\% with a step of 2\%), respectively. We then estimate population-weighted concentration changes and calculate avoided deaths in each scenario based on most recent concentration-response functions. Although the amount of coal use in the industrial sector is about 40 times higher than rural and residential coal use, the deaths avoided by reducing industrial coal use are only about 4 times higher, suggesting unit pollutant emissions and marginal damage of rural and residential coal use are substantially higher. This result is consistent with the recent finding by Yun et al.{\cite{Yun2020}} that China's residential sector contributed only 7.5\% of energy consumption but contributed 27\% of primary PM$_{2.5}$ emissions and 23\% of the outdoor PM$_{2.5}$ concentrations, respectively. To showcase the computational advantage of our framework, we further downscale the health benefit calculation by estimating marginal damages of fossil fuel use (per ton of coal equivalent) in these sectors by each grid cell (10 $km$ $\times$ 10 $km$). As expected, there is a wide spread of marginal damage of fossil fuel consumption across different sector and also within each sector spatially. We find that reducing rural and residential coal use offers the highest health benefits per ton of use reduction. Further information about the framework is provided in Methods. 

\vspace{5mm} %5mm vertical space
\noindent \textbf{Health co-benefits estimations} 

\begin{figure*}[ht!]
\centerline{\includegraphics[width=1\textwidth]{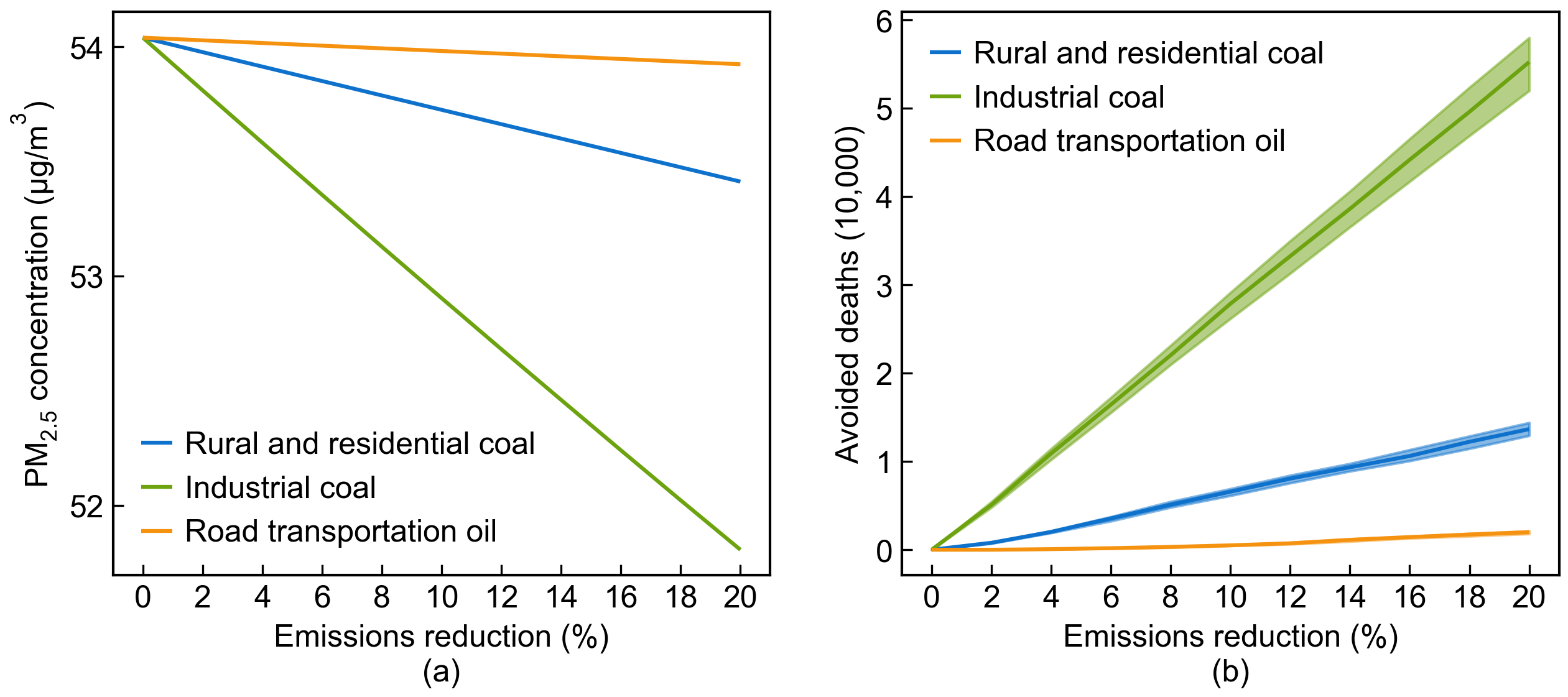}}
\caption{China's population-weighted PM$_{2.5}$ concentrations and corresponding avoided deaths (shaded areas represent 90\% confidence intervals) in 2015 if fossil energy use in a polluting sector were curtailed by a certain percentage (from 2\% to 20\%).}
\label{fig:fig2}
\end{figure*}

We first illustrate how our model could be applied in conventional scenario analyses to estimate air quality and health benefits changes with policies aiming to reduce fossil fuel use. We select three key combustion sources in China as our targets, i.e., industrial coal use, rural and residential coal use, and road transportation oil use. \textbf{Figure \ref{fig:fig2}} shows the national population-weighted PM$_{2.5}$ concentration and corresponding avoided premature deaths if fossil energy use in one aforementioned sector is curtailed by a certain percentage (from 2\% to 20\% with a step of 2\%), holding other emissions sources and meteorological conditions unchanged. We find the PM$_{2.5}$ concentration decreases and corresponding avoided deaths increase almost linearly with the size of fossil energy curtailment. Annual avoided deaths from reducing 20\% of fossil fuels from these three sectors are 55 thousands, 14 thousands, and 3 thousands for industrial coal, rural and residential coal, and road transportation oil, respectively. Avoided deaths from reducing industrial coal and rural and residential coal use have the same order of magnitude, although CO$_2$ emissions (and energy consumption) from industrial coal use are one order of magnitude higher than emissions from the other two sectors (7.4 gigatons for industrial coal, 0.2 gigatons for rural and residential coal, and 1.0 gigaton for road transportation oil). This result reflects the fact that emissions factor of industrial coal use is significantly lower thanks to strict end-of-pipe measures implemented in China recently, and that locations of emissions sources might be further away from densely-populated areas. While most effective air quality control measures have been targeted at the industrial sector over the past few years{\cite{Zhang2019}}, there are enormous cost-effective potentials to further harvest health benefits from reducing rural and residential coal use.

\begin{figure*}[ht!]
\centerline{\includegraphics[width=1.1\textwidth]{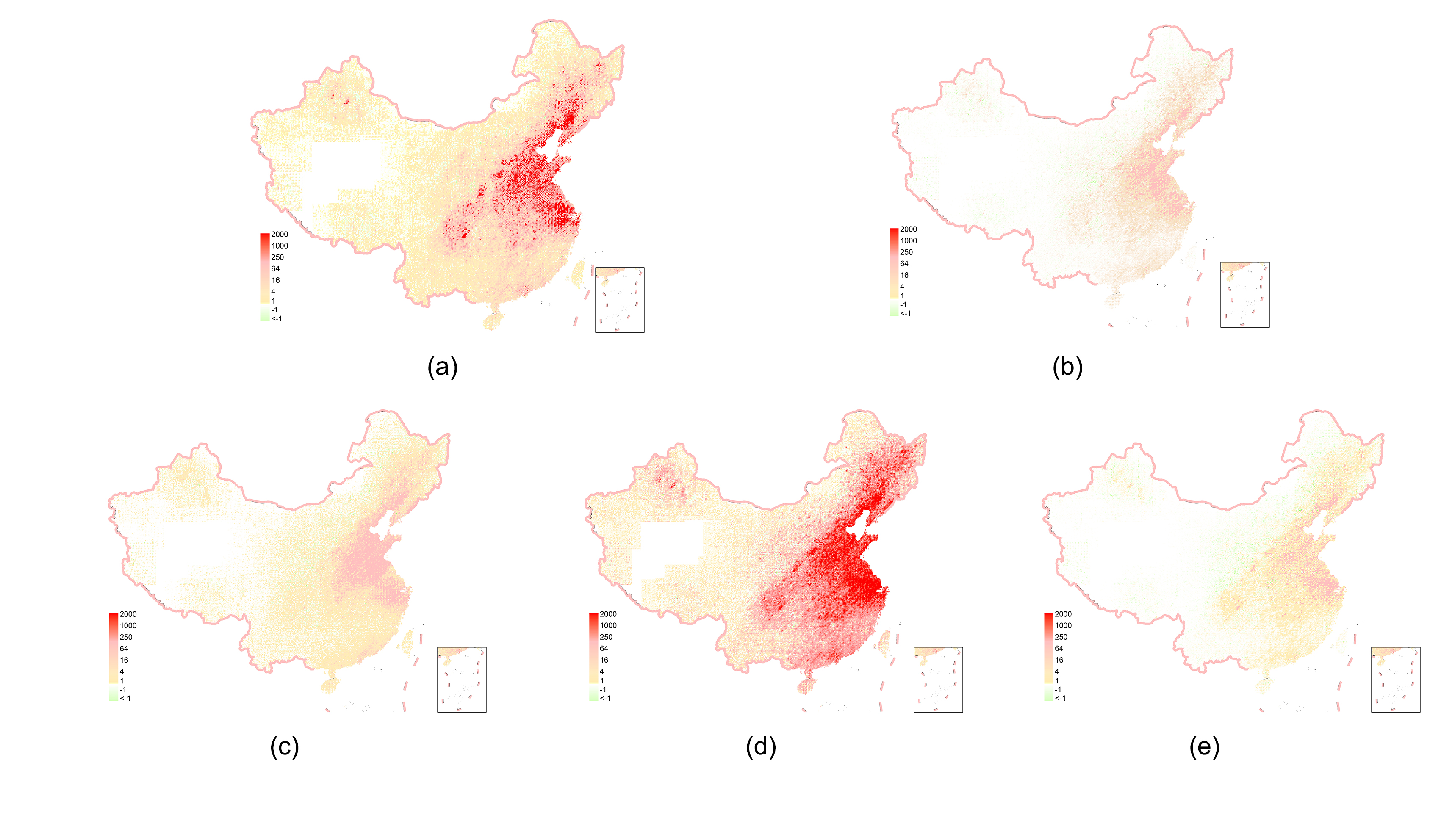}}
\caption{Marginal damages measured in dollars attributable to an additional ton of CO$_2$ emissions from (a) rural and residential coal use, (b) coal use in the industry sector, (c) oil use in the industry sector, (d) coal use in the service sector, and (e) oil use in the transportation sector.}
\label{fig:fig3}
\end{figure*}

We then demonstrate how our model could offer estimated spatial distribution of marginal damage of fossil energy use in different sectors, a task that will require unacceptably long simulation time using conventional CTMs. We show marginal damages attributable to an additional ton of CO$_2$ emissions from a certain type of fossil energy use at every source location in mainland China in \textbf{Figure \ref{fig:fig3}}, with maps for five major air-polluting emissions sources considered in our model (i.e., industrial coal use, rural and residential coal use, coal use in the service sector, road transportation oil use, and industrial oil use). We adopt the value of statistical life (VSL) estimate as 1.8 million in US\$(2015), see Methods for details. Average marginal damage per unit ton of CO$_2$ emissions ranges from 17 to 360 dollars (US\$ 2015 price) for these sectors, with highest damages for emissions from rural and residential coal use (360 \$/ton) followed by coal use in the service sector (213 \$/ton). Damages for emissions from industrial coal use, road transportation oil and industrial oil use are one order of magnitude lower (23 \$/ton for industrial coal use, 23 \$/ton for industrial oil use, and 17 \$/ton for road transportation oil use). The mean and standard deviation of marginal damages attributable to an additional ton of CO$_2$ emissions in China are 34 \$/ton and 84 \$/ton, respectively. The mean value has the same order of magnitude compared to the estimate with constant VSL of 1.5 million US\$(2005) in Vandyck et al.{\cite{Vandyck2018}}. Similar to the pattern found in Goodkind et al.{\cite{Goodkind2019}}, all the distributions are positively skewed, indicating high marginal damages for some hotpots, especially for rural and residential coal use (see \textbf{Supplementary Figure 2}). The huge spread across energy use sources and locations reflect substantial differences in coal quality, combustion condition, and end-of-pipe treatment. Overall, the mean marginal damage of CO$_2$ emissions from coal and oil use in China is 37 and 19 \$/ton, respectively.

\begin{figure*}[ht!]
     \centering
\begin{subfigure}[h]{0.5\textwidth}
     \centering
     \includegraphics[width=\textwidth]{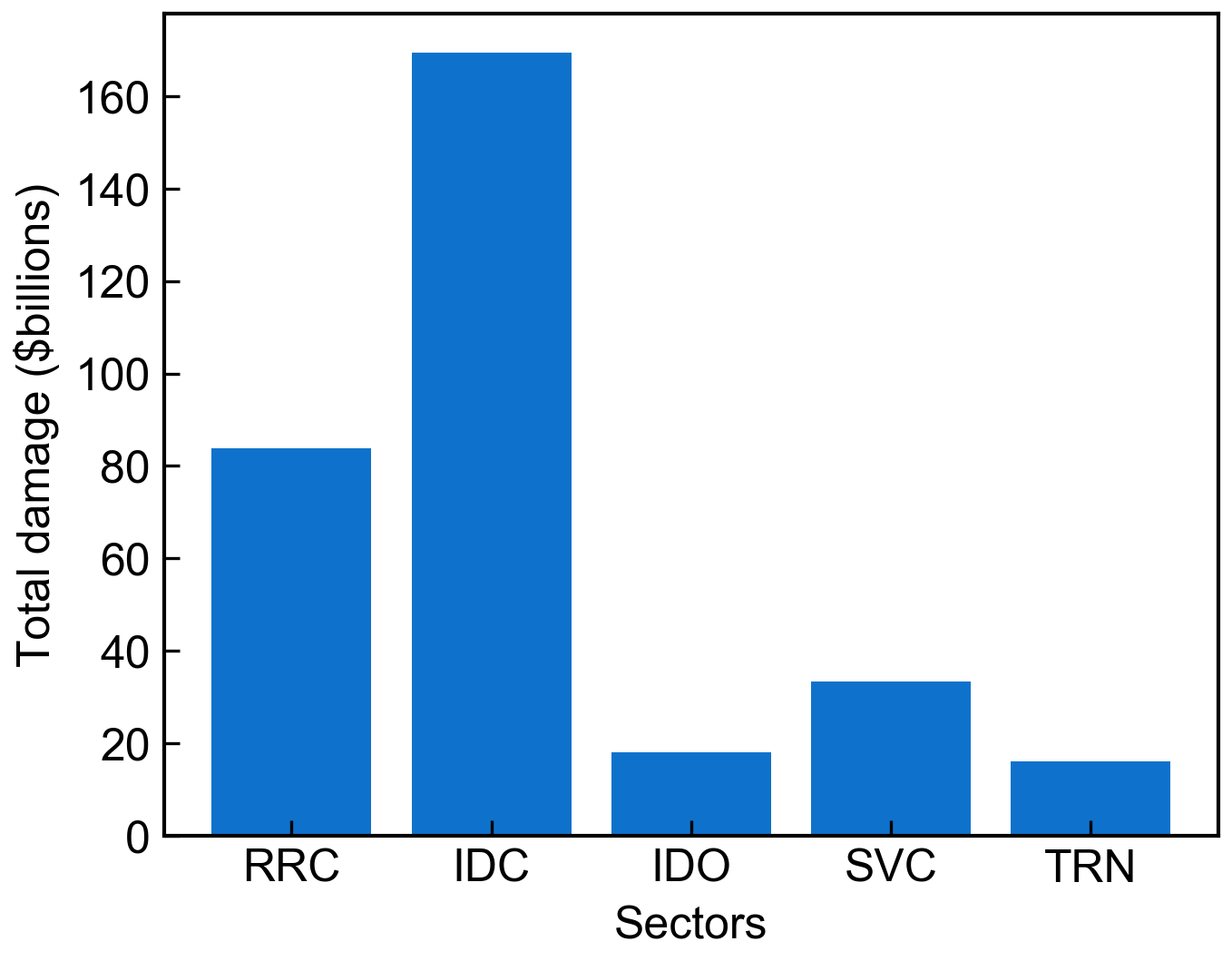}
     \caption{}
 \end{subfigure}
     \vfill
 \begin{subfigure}[h]{1\textwidth}
     \centering
     \includegraphics[width=\textwidth]{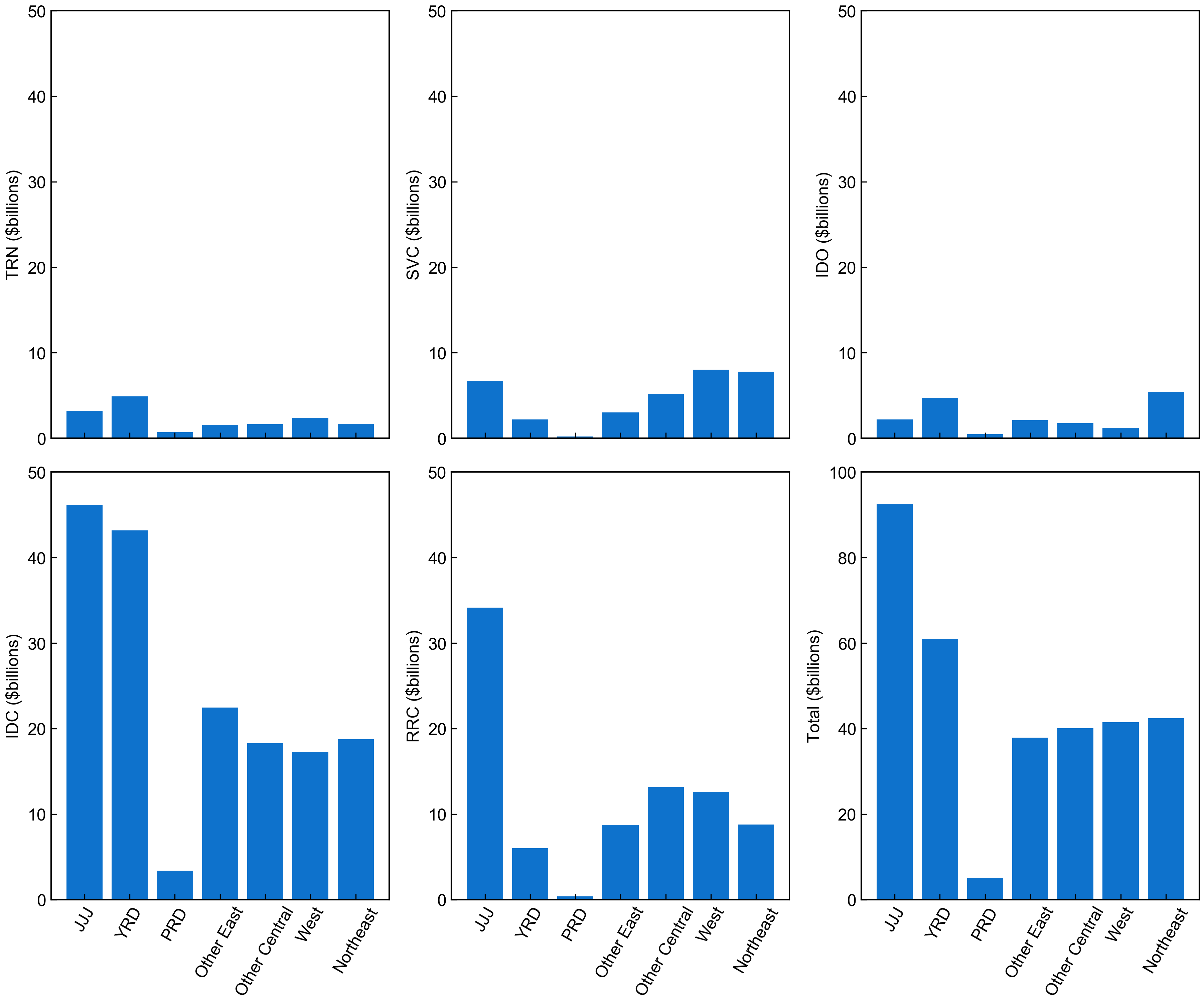}
     \caption{}
 \end{subfigure}
\caption{Total damages from fossil fuel use in China (a) by sector and (b) by region with sectoral breakouts.}
\label{fig:fig4}
\end{figure*}

Combining estimated marginal damages and fossil energy use data with sectoral and regional source information, we can calculate total damages by sector and by region. For illustration purposes, we aggregate all the mainland Chinese cities into seven regions following the literature {\cite{Liang2019,Zhang2019}} and China's official document{\cite{NBS2011}}: Bejing-Tianjin-Hebei (JJJ representing first characters of three provinces' short names) and surrounding cities (some cities in Shanxi, Shandong, and Henan province included), Yangtze River Delta provinces (YRD), Pearl River Delta provinces (PRD), other East, other Central, West, and Northeast (the first three regions are China's major air-pollution control regions). \textbf{Figure \ref{fig:fig4}} (a) shows total damages by sector. Industrial coal use has the largest total damage (around \$160 billion), while the total damage of rural and residential coal use is nearly one half of the industrial coal's total damage. The other types of fossil fuel use have one order of magnitude smaller total damages. \textbf{Figure \ref{fig:fig4}} (b) shows total damages with sectoral breakouts by region. As one of the most populous and polluted region, the Bejing-Tianjin-Hebei and surrounding cities bear the highest total damage from fossil fuel emissions (around \$90 billion). Its rural and residential coal use imposes substantially higher total damages than other regions because winter heating contributes much more to the total PM$_{2.5}$ pollution compared to other major air-pollution control regions with less heating demand, e.g., Yangtze River Delta where the total damage is the second highest (around \$60 billion). Some regions show high damages from a specific type of fossil energy use, e.g., industrial oil use in the Northeast, reflecting the relatively concentrated energy use of this type or associated high marginal damage in some hot spots in these regions.

% ======================================================================
\section{Robustness analysis}
% ======================================================================

We examine the robustness of emissions-concentration relationship derived from our approach by re-calculating the average marginal damage using models trained by different sets of hyper-parameters. We select four additional sets of hyper-parameters that produce smallest weighted least square errors on the validation data sets after the set of hyper-parameters chosen for our base-case model. The mean and standard deviation of marginal damage of CO$_2$ is 35--40 \$/ton and 61--100 \$/ton, which are close to the results from our base-case model (34 \$/ton and 84 \$/ton).

We then evaluate if the selection of surrounding area size for the PM$_{2.5}$ monitoring station could affect the prediction accuracy our framework. Ideally, we should include an area as large as possible to incorporate the impacts of possible long-distance pollutant transportation. In practice, certain thresholds are usually chosen to avoid data availability and computational issues. Besides our base-case model with input tensors representing 610 $km$ $\times$ 610 $km$ area centered by a monitor station, we report the prediction accuracy of two alternative models with smaller input tensors (210 $km$ $\times$ 210 $km$ and 410 $km$ $\times$ 410 $km$) but identical training process. Surprisingly, models with smaller input tensors can achieve almost similar accuracy compared to our base-case model on the training data set. Our base-case model shows better accuracy on the validation data set, but no significant supremacy on the test data set, see \textbf{Supplementary Figure 3}. This result suggests that emissions from a closer surrounding area (210 $km$ $\times$ 210 $km$) dictate the concentration prediction results, and our modeling framework could achieve satisfactory accuracy as long as most relevant input data (fossil energy consumption within a certain distance from the station) are included.

\begin{figure*}[ht!]
\centerline{\includegraphics[width=0.6\textwidth]{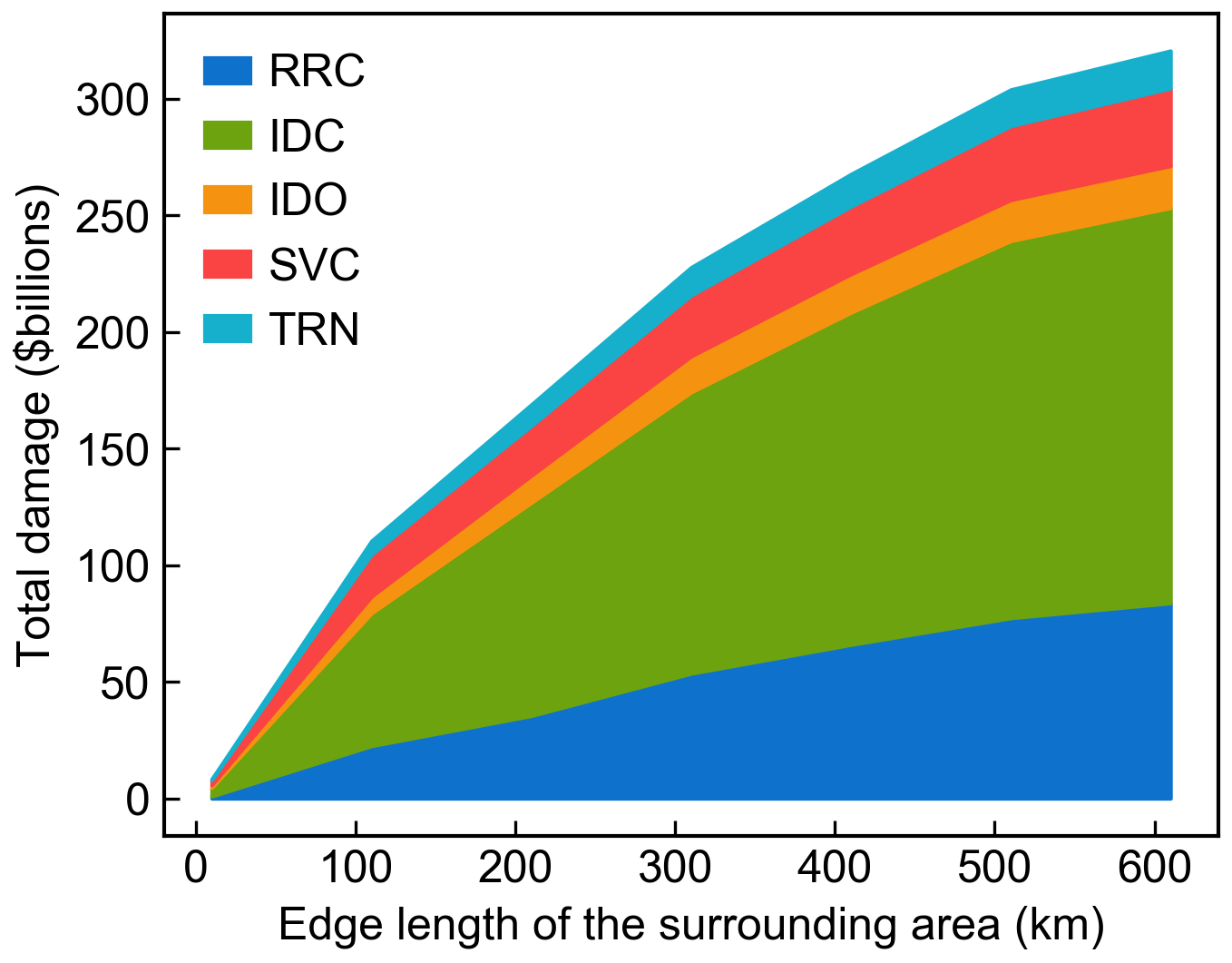}}
\caption{Total damage by distance (measured by the edge length of the square centered at a monitor station) and emissions source type.}
\label{fig:fig5}
\end{figure*}

To further illustrate the above finding, we calculate the total damage within different sizes of surrounding area using our base-case model. \textbf{Figure \ref{fig:fig5}} shows the total damage caused by emissions within a certain size of square (from 110 $km$ $\times$ 110 $km$ to 610 $km$ $\times$ 610 $km$) centered by a monitor station. The total damage curve has a concave relationship with regards to the surrounding area size. Emissions from a faraway source has a smaller impact on the PM$_{2.5}$ concentration of the destination. More than half of the total damage caused by emissions from the full 610 $km$ $\times$ 610 $km$ area occur within the centering 210 $km$ $\times$ 210 $km$ area (one ninth of the full area). Compared to the results found by Goodkind et al.{\cite{Goodkind2019}} (half of total PM$_{2.5}$ damages within a 4,096 km radius circle are incurred by people living within 32 km of a source), the total damage curve in China is less concave, possibly because different distributions of pollution sources and population centers that allow the atmospheric transmission to form more local pollution from faraway sources in China.

% ======================================================================
\section{Discussion}
% ======================================================================

It is well-recognized that marginal damage of emissions and corresponding health benefits of abatement vary widely by place and pollution type. Traditional complicated chemical transport models are not suitable for economic and policy analysis that requires many model iterations to explore the variation of marginal damages. Recent developments of reduced-form models with satisfactory approximations for full-scale models have allowed the researchers to quantify the heterogeneity of the marginal damage but still place some limitations for broader applications. In this study, we develop a machine-learning based framework that can produce more accurate, accessible, and easy-to-update simulations for air quality, or more specifically, PM$_{2.5}$ concentrations. Applying this framework to estimate the marginal damage of fossil fuel CO$_2$ emissions by fuel type and source location in China provides us with important insights into future policy designs that aim to achieve ambitious air quality and climate targets.

We find that China's current carbon pricing stringency does not match the magnitude of marginal damage of per ton of CO$_2$ emissions. Although many command-and-control climate policies are in place in China, limited market-based instruments have been applied. Currently, only seven regional CO$_2$ emissions trading schemes (ETS) are in operation with highest average carbon price at the magnitude of \$15/ton in Beijing pilot ETS{\cite{Worldbank}}, substantially less than the average marginal damage (\$/ton) estimated by this study. We urge a more accelerated development of China's national ETS to form an appropriate carbon price that can better internalize the public health externality of CO$_2$ emissions.

The highly heterogeneous marginal damage by fuel type and location has also important implications. By far, only CO$_2$ emissions from industrial fossil fuel use are planned to be covered by China's national ETS{\cite{PizerZhang2018}}. Our analysis has shown, however, CO$_2$ emissions from industrial fossil fuel have a much lower marginal damage compared to the emissions from coal use in the rural and residential as well as service sector. Therefore, complementary policies that place an effective carbon price on these sectors to encourage fuel switching (e.g., coal to natural gas for small boilers or electrification of residential heating) are essential. We also note that the marginal damage of CO$_2$ emissions in more populous eastern China, particularly in the North China Plain and Yangtze River Delta region, is much higher than that of emissions in the rest of the country, suggesting a trading ratio{\cite{HollandYates2015}} could be considered in the future design of China's national ETS.

Applying machine learning techniques in the environmental integrated assessment motivates fruitful future work. Although our framework provides satisfactory results for the annual PM$_{2.5}$ concentration estimation, prediction accuracy for pollutants other than particular matters needs further improvement. \textbf{Supplementary Figure 4} shows the prediction accuracy for PM$_{10}$, SO$_2$, NO$_2$, CO, and O$_3$ by applying the same framework and training process to these pollutants. Accuracy of PM$_{10}$ is similar to that of PM$_{2.5}$. For SO$_2$, NO$_2$, and CO, the framework can achieve acceptable fitting results on the training data set but fail to achieve consistent accuracy on the validation and test data sets; the fitness is low even on the training data set for ozone. Refining the framework architecture and training process and incorporating relevant atmospheric condition data for pollutants other than particular matters remains an interesting direction for future research.

We believe it is important to extend the co-benefits analysis implemented in our study to more developing countries with imperative air pollution control and climate mitigation policy development. High-resolution marginal damage estimations that are very relevant for policy making have been almost exclusively conducted in the United States {\cite{MullerMendelsohn2009,Dedoussi2020,Holland2016,Tessum2019,Goodkind2019,Hill2019,Mayfield2019}}, suggesting the complexity of the methodology even with elaborately-designed reduced-form CTMs. We hope our machine learning framework would supply a more accessible tool for researchers and policymakers to implement more comprehensive and timely analysis in more developing countries that are constrained with emissions inventory and atmospheric measurement data.

%TC:ignore
% ======================================================================
\section{Methods}
% ======================================================================
Methods, including statements of data availability and references, are available in the online version of this paper.

% ======================================================================
% \section{References}
% ======================================================================

\newpage
\noindent
Correspondence and requests for materials should be addressed to Shenhao Wang and Bofeng Cai (E-mail: shenhao@mit.edu and caibf@caep.org.cn).

% ======================================================================
\section{Acknowledgments}
% ======================================================================
We acknowledge the support of the National Science Foundation of China (Project No. 71690244). Da Zhang has been supported by the MIT Joint Program on the Science and Policy of Global Change, funded through a consortium of industrial sponsors and Federal grants, including the U.S. Department of Energy (DOE) under Integrated Assessment Grant (DE-FG02-94ER61937). We thank Jiacheng Cui and Xinhao Wang for their excellent research assistance. 

% ======================================================================
\section{Author contributions}
% ======================================================================
D.Z., B.C., and S.W. conceived the research. D.Z., Q.W., and S.W. performed the modeling simulations. All authors discussed the results and contributed to the writing of the paper.

% ======================================================================
\section{Competing financial interests}
% ======================================================================
The authors declare no competing financial interests.
%TC:endignore

%TC:break _Methods_
% ======================================================================
\section{Methods}
% ======================================================================

\textbf{Data sets and preprocessing.} 
We denote the inputs used to predict annual average fine particle (PM$_{2.5}$) concentration as $x_{ijkn} \in \mathbb{R}^4$ with four indices: $n$ is the index of air-quality monitoring stations ($ 1 \leq n \leq N$); $k$ is the index of different types of data inputs that are used for the prediction, including the level of fossil energy use of a specific energy type in a specific sector as well as some additional geographic information, for example, altitude, temperature, and precipitation ($ 1 \leq k \leq K$); and $(i,j)$ is the location index representing a grid cell of an area centered around the station ($1 \leq i,j \leq I, J$). Therefore, $x_{ijkn}$ represents the value of a specific input $k$ for a grid cell  $(i,j)$ that is close to a specific station $n$. The total number of stations is $N = 943$. We include eight ($K=8$) types of data inputs for each station in the form of 2D tensors ($I,J$). Specifically, these eight types of data inputs are: (1) rural and residential coal use (RRC), (2) industrial coal use (IDC), (3) industrial oil use (IDO), (4) coal use in service industry (SVC), (5) oil use in road transportation (TRN), (6) altitude (ALT), (7) annual average temperature (TEM), and (8) annual average precipitation (PCP). The number of pixels in each 2D tensor in our main model specification is $I \times J = 61 \times 61$. With each pixel representing a $10 \ km \times 10 \ km$ cell, the geographical size of each 2D tensor equals to $[I \times 10] \times [J \times 10] = 610 \ km \times 610 \ km $. The large geographical coverage of each station, along with $943$ stations in total, guarantees that our analysis covers all the territory of mainland China with permanent residents. 

The output of our main model is the annual average PM$_{2.5}$ concentration, denoted as $y_n$. Note that each $y_n \in \mathbb{R}$ is a real number for each station, different from the high dimensionality of the input variable $x_{ijkn}$. We also train separate models to predict annual average concentrations of other air pollutants, including the annual average concentration of PM$_{2.5}$, PM$_{10}$, SO$_2$, NO$_2$, $CO$, and O$_3$. While the output of the models are different, the model structure and training process are similar to our main model specification.

Inputs $x_{ijkn}$ are normalized before modeling to improve prediction accuracy and facilitate training process{\cite{Hinton2012}}. The input $x_{ijkn}$ is normalized as following:

\begin{equation}
x_{ijkn} \leftarrow \frac{x_{ijkn} - \bar{x}_{kn}}{\sigma_{kn}}
\end{equation}

in which

\begin{equation}
 \bar{x}_{kn} = \frac{1}{IJ} \underset{i=1,j=1}{\sum^{I,J}} x_{ijkn}; \ \ \ 
 \sigma_{kn} = \sqrt{\frac{1}{IJ} \underset{i=1,j=1}{\sum^{I, J}} ( x_{ijkn} - \bar{x}_{kn} )^2} 
\end{equation}

For stations that are close to national boundaries or sea, the 2D tensors could have cells that we do not have energy use information. We fill these cells with zeros.

\bigskip
\noindent\textbf{ResCNN model architecture.}
As visualized in \textbf{Supplementary Figure 1}, our model architecture consists of two parts. The upper part is a standard AlexNet{\cite{Krizhevsky2012}}, which uses the normalized $3D$ tensor $x_{ijkn}$ as input for each station $n$. It starts with repeated blocks of convolutional layers and max pooling layers, and ends with several fully connected layers. The bottom part is similar to linear regression that uses linear specification and average value of each $1D$ tensor $\bar{x}_{kn}$ as input. The two parts provide complementary information: AlexNet absorbs the nonlinear structural spatial information of each $3D$ tensor, while the linear part absorbs the magnitude information of each $3D$ tensor, which is lost in AlexNet due to normalization. Intuitively, PM$_{2.5}$ concentration varies with the average amount of the energy use (e.g., higher fossil energy use usually causes higher pollution concentration), and it also depends on some spatial structural information which is missing in the average values (e.g., fossil energy use that is closer to the station will have a larger effect on the concentration, and the spatial distribution of one type of fossil energy use may have an nonlinear interaction with the spatial distribution of another type of fossil energy use). Different from the traditional process of explicitly specifying the nonlinear relationship, our model automatically learns the nonlinear relationship between energy consumption, atmospheric conditions, geographic information, and resulting annual average concentration level of PM$_{2.5}$. The synthesis of the AlexNet and the linear parts is similar to ResNet {\cite{HeKaiming2016}}, which is also a linear combination of linear and nonlinear feature maps. We therefore name the model architecture as ResCNN because it combines linear regression to capture the magnitude effect and CNN to capture the residual nonlinear relationship. Mathematically, the ResCNN model is

\begin{equation}
y_n = f(x) = f_c (x_{ijkn}) + f_l(\bar{x}_{kn}) = (g_L \circ g_{L-1} \circ ... \circ g_1)(x_{ijkn}) + w_{K}' \bar{x}_{kn}
\label{eq:cnn}
\end{equation}

In Equation \ref{eq:cnn}, $(g_L \circ g_{L-1} \circ ... \circ g_1)(x_{ijkn})$ represents the AlexNet architecture with $L$ layers; $w_{K}' \bar{x}_{kn}$ represents the linear part. % Corresponding to convolutional layer, max pooling layer, and fully connected layer, $g_l$ takes the following three forms:

% \begin{flalign}
% & g_l(i,j) = max \{ \sum_{p=1,q=1}^{P^c,Q^c} g_{l-1}(i + p, j + q) w_{p,q}, 0 \} \\
% & g_l(i,j) = \underset{ \substack{p \in \{1,2...,P^m \}, \\ q \in \{1,2...,Q^m \}} }{max} \{ g_{l-1}(i+p, j+q) \} \\
% & g_l(m) = w'_{lm} g_{l-1}
% \label{eq:cnn_eq}
% \end{flalign}

% in which $P^c$ and $Q^c$ are the size of convolutional kernels; $P^m$ and $Q^m$ are the size of max pooling kernels; $w_{lm}$ represents one vector of coefficients at layer $l$. One challenge in CNN modeling is how to specify the values of $P^c$, $Q^c$, $P^m$, $Q^m$, and many other hyper-parameters, such as depth and width of CNN, which dictate model performance. 

% \bigskip
% \noindent\textbf{Linear model.}
% To set a reference for the performance of our ResCNN model, we construct a linear regression model for prediction. The linear regression uses $\bar x_{kn} $ of each industrial section in each station as the input. In fact, it is the same as the bottom part of the ResCNN model in \textbf{Supplementary Figure 1}:

% \begin{equation}
% y_n = f_{l}(\bar x_{kn}) = w_{K}^T \bar x_{kn} 
% \end{equation}

% This alternative linear model is therefore one subset of the ResCNN model, so it should not be a surprise that the ResCNN could outperform the linear model. Put it differently, the ResCNN model could be treated as an enhanced linear model with AlexNet capturing nonlinear structural information. By comparing the ResCNN to linear regression, we could demonstrate the additional explanation power of the automated feature learning in the ResCNN architecture to predict PM$_{2.5}$ concentration.

\bigskip
\noindent\textbf{Evaluation Metrics.}
To evaluate the performance of the ResCNN, we have adopted some common evaluation metrics for the performance of the CTMs, for example, coefficient of determination ($R^2$), normalized mean bias (NMB), and normalized mean error/mean absolute percentage error (NME/MAPE), from the literature{\cite{Muller2011,Tessum2017,VanDingenen2018,Appel2012,Zhong2016}}. We add an important modification to all of the evaluation metrics by introducing normalized weights $w_n$ that are proportional to the population that a specific station $n$ corresponds to. For example, if there is only one station that enters our sample for a specific city, we assume the station generates the concentration reading that represents the exposure of the city's whole population; however, if multiple stations are present in our sample for a city, we assume each station generates the concentration reading that represents an equal share of the city's population. With this modification, the performance measure gives more weights on the prediction accuracy of stations corresponding to larger population because the marginal damage we estimate later is in proportional to the population. We illustrate all the metrics we compute in \textbf{Supplementary Table 1}. Our main model uses the weighted least square errors on the validation set in hyper-parameter searching, which we discuss in the section below.

\bigskip
\noindent\textbf{Hyper-Parameter searching and training.}
To address the challenge of specifying the appropriate hyper-parameters, this study combines random search and grid search. First, random searches (400 trials) in a pre-specified, larger hyper-parameter space are done. Conditioning on each hyper-parameter and while keeping the others random, hyper-parameter values that yield consistently lower performance or higher variance in performance are pruned. Then a complete grid search was done to the remaining hyper-parameter values (384 combinations). 
While numerous methods can be used to identify the best hyper-parameter, the simple random search{\cite{Bergstra2012}} is still a useful benchmark, even in comparison to more complicated hyper-parameter searching methods based on reinforcement learning or Gaussian process {\cite{Zoph2016,Zoph2017,Snoek2015}}. The full hyper-parameter space is shown in \textbf{Supplementary Table 2}, and the pruned hyper-parameter space is shown in \textbf{Supplementary Table 3}. Among the hyper-parameters, some are model specific and require some searching in each specific application. For instance, dropout and batch normalization were found as effective regularization in several studies {\cite{Srivastava_Hinton2014,Ioffe2015}}. Data augmentation {\cite{Goodfellow2016}} assumes the invariance property: when images are rotated or flipped, the new images should not change the predicted values $\hat y$. On the other hand, some hyper-parameters are set to follow the common practice. For instance, Rectified Linear Unit (ReLU) is used as activation functions for each neuron; He initialization{\cite{Geron2017}} is used to address the problem of vanishing and exploding gradients; Adam optimizer{\cite{Kingma2014}} is used for gradient descent optimization. 

The ResCNN model is trained by empirical risk minimization (ERM). Formally,

\begin{equation}
 \underset{W}{min} \ E(W; W_h) = \underset{W}{min} \frac{1}{N} \sum_{n=1}^{N} w_n \times (y_n - f(x_{ijkn}; W, W_h))^2 
\label{eq:cnn_training}
\end{equation}

in which $W$ represents parameters and $W_h$ represents hyper-parameters, while $w_n$ represents the weight of each observation as we describe in the above section. Note that the training in Equation \ref{eq:cnn_training} is conditioning on the specific choice of hyper-parameters $W_h$. Denote $W^* = \underset{W}{argmin} \ E(W; W_h)$, the optimum hyper-parameter $W_h^*$ is chosen by random searching:

\begin{equation}
W_h^* = \underset{W_h \in W_h^{(1)}, W_h^{(2)}, ... , W_h^{(S)}}{argmin} E(W^*; W_h)
\label{eq:cnn_hyper_training}
\end{equation}

$W_h^{s}$ represents random sample from the hyper-parameter space. The best hyper-parameter $W_h^*$ is chosen out of $S=100$ training. To train model and choose hyper-parameters, the full dataset is split into training, validation, and testing sets with the ratio equals to $3:1:1$. The training set is used to train ResCNN model as in Equation \ref{eq:cnn_training}; validation set is used for the selection of hyper-parameter as in Equation \ref{eq:cnn_hyper_training}; testing set is used for model evaluation and comparison.

\bigskip
\noindent\textbf{Health impacts and valuation}
We apply the up-to-date GEMM NCD+LRI method\citep{Burnett2018} to estimate avoided premature death related to reductions in chronic exposure to outdoor fine particulate matter (PM$_{2.5}$) under different scenarios. The GEMM NCD+LRI method is considered as a major update to the widely-used IER 5-COD approaches\citep{Burnett2014,Cohen2017} and has been adopted in some recent cost-benefit analysis\citep{Dedoussi2020,Liang2019,Zhang2019}. Compared to the IER approach, the GEMM NCD+LRI method incorporates findings from recent cohort studies that quantify the relationship between avoided premature death and ambient PM$_{2.5}$ concentration (C-R relationship) in regions with high PM$_{2.5}$ concentration levels, making it more appropriate to be applied in relatively more polluted countries like China. In addition, the GEMM NCD+LRI method associates PM$_{2.5}$-related death to non-accidental deaths caused by noncommunicable diseases and lower respiratory inflections, a more comprehensive range than the five specific causes of death considered in the IER 5-COD approaches.

The GEMM NCD+LRI quantifies the relationship between hazard ratio ($RR$) of NCD+LRI and ambient PM$_{2.5}$ concentration ($c$) with the following equation:

\begin{equation}
\begin{split}
RR(c) = exp(\theta\times\frac{\ln(\frac{max(0,c-c_f)}{\alpha}+1)}{1+exp(-\frac{max(0,c-c_f)-\mu}{\upsilon})})
\end{split}
\end{equation}

\noindent
where $\theta$, $\alpha$, $\mu$, $\upsilon$ and $c_f$ are all shape parameters that define the C-R relationship. Since the baseline mortality rate is different for adults with different ages, we follow the convention to divide population in a specific grid cell $n$ into 12 subgroups (adults with age from 25 to 85 and above in five-year intervals). Consider two scenarios (scenario 0 and scenario 1) with different ambient PM$_{2.5}$ concentration levels ($c_0, c_1$), the avoided death under scenario 1 compared to scenario 0 in grid cell $n$ is calculated by summing up avoided deaths of all $m$ age groups:

\begin{equation}
\label{eqn:delta_Mn}
\begin{split}
\Delta M_n = \sum_m M_m^B \times pop_{m,n}\times(\frac{1}{RR(c_{1,n})}-\frac{1}{RR(c_{0,n})})
\end{split}
\end{equation}

\noindent
where $M_m^B$ is the baseline mortality rate for age group $m$ in China, retrieved from the Global Health Data Exchange, and $pop_{m,n}$ is the population of age group $m$ in grid cell $n$.

We adopt the assumption that $\theta$ has a normal distribution with mean $\mu_{\theta}$ and standard deviation $\sigma_{\theta}$\citep{Burnett2014}. We can then sample 1,000 points from the normal distribution and calculate the mean and 95\% confidence interval of avoided death using Equation \ref{eqn:delta_Mn}. An alternative faster approach to obtain the mean avoided death $\overline {\Delta M_n}$ is to use the following equation to first calculate the mean of $1/RR(c_{n})$, which has a lognormal distribution,

\begin{equation}
\begin{split}
\overline{1/RR(c_{n})}=exp(-\mu_\theta \frac{\ln(\frac{max(0,c_n-c_f)}{\alpha}+1)}{1+exp(-\frac{max(0,c_n-c_f)-\mu}{\upsilon})}+\frac{1}{2}({\sigma_\theta \frac{\ln(\frac{max(0,c_n-c_f)}{\alpha}+1)}{1+exp(-\frac{max(0,c_n-c_f)-\mu}{\upsilon})}})^2)
\end{split}
\end{equation}

\noindent
and then calculate the mean avoided death $\overline {\Delta M_n}$.

\begin{equation}
\label{eqn:mean_delta_Mn}
\begin{split}
\overline {\Delta M_n} = \sum_m M_m^B \times pop_{m,n}\times \left( \overline{1/RR(c_{1,n})}-\overline{1/RR(c_{0,n})} \right)
\end{split}
\end{equation}

In our scenario analysis, where sector $k^{\prime}$ emission is curtailed by $p$ ($p$ ranges from $2\%$ to $20\%$), $c_{0,n}$ is taken to be the predicted baseline PM$_{2.5}$ concentration, and $c_{1,n}$ is obtained by reducing all input emissions in sector $k^{\prime}$ by $p$:
\begin{equation}
    c_{1,n} = f(x_{ijkn, k\neq k^{\prime}}, x_{ijk^{\prime} n} * (1-p); W^*, W_h^*))
\end{equation}

The marginal change of PM$_{2.5}$ due to the marginal change of emissions can be obtained from the gradients of PM$_{2.5}$ with respect to the input emissions from each sector and each grid cell ($\frac{\partial y}{\partial x_{ijk}}$). Since back propagation is used for training the neural network, the gradients can be directly exported from the models. For each unit emissions increase in sector $k ^{\prime}$ in cell $i,j$, the mean avoided death for station $n$ is:

\begin{equation}
\overline {\Delta M_{ijk^{\prime} n}}  = \sum_m M_m^B \times  pop_{m,n} \times  \left[ \overline{1/RR \left(c_{0,n} + \frac{\partial y_n}{\partial x_{ijk^{\prime} n}} \right)}-\overline{1/RR(c_{0,n})} \right]
\end{equation}

\noindent where again $c_{0,n}$ is taken to be the predicted baseline PM$_{2.5}$ concentration.

We close the estimate of marginal monetary damages related to increased premature mortality due to an additional ton of CO$_2$ emissions by using an inferred value of statistical life (VSL) for China based on the U.S. EPA recommended VSL of 8.7 million in US\$(2015){\cite{EPA2010}}. We adopt the income elasticity for high-income countries recommended by a recent meta-study{\cite{Robinson2019}} for extrapolating from the U.S. VSL ($VSL_{base}$) to the VSL for China ($VSL$):

\begin{equation}\label{eqn:eqVSL}
VSL = VSL_{base} \times (pcGDP_{China}/pcGDP_{U.S.})^{0.8}
\end{equation}

\noindent where pcGDP$_{China}$ and pcGDP$_{U.S.}$ represent GDP per capita in 2015 for China and the U.S., respectively. Equation \ref{eqn:eqVSL} gives the VSL estimate as 1.8 million in US\$(2015) for China.

Combining the considerations of premature mortality and the inferred value of statistical life, the marginal monetary damages due to an additional ton of CO$_2$ emissions in sector $k ^{\prime}$ in cell $i,j$ is 

\begin{equation}
    MD_{ijk^{\prime}} =  - \sum_{n} VSL \times \overline {\Delta M_{ijk^{\prime} n}} 
\end{equation}

The total monetary damages incurred by emissions from sector $k^{\prime}$ are the sum of the production of marginal monetary damages and emissions in each cell:

\begin{equation}
    TD_{k^{\prime}} = \sum_{i,j}{MD_{ijk^{\prime}} \times x_{ijk^{\prime}}}
\end{equation}

%TC:ignore
\bigskip
\noindent\textbf{Data availability} The scripts and data that support the findings of this study are available online (web address to be announced).

% ======================================================================
% \section{References}
% ======================================================================

\makeatletter
\apptocmd{\thebibliography}{\global\c@NAT@ctr 34\relax}{}{}
\makeatother

\section{Figure Captions}

\textbf{Figure 1} Comparison of model-predicted and observed PM$_{2.5}$ concentrations at 943 stations in mainland China for year 2015 by training, validation, and test data sets.
%Model prediction performance metrics include mean bias (MB), mean error (ME), mean fractional bias (MFB), mean fractional error (MFE),  mean absolute percentage error (MAPE), root mean square error (RMSE), and $R^2$.

\noindent\textbf{Figure 2} China’s population-weighted PM$_{2.5}$ concentrations and corresponding avoided deaths (shaded areas represent 95\% confidence intervals) in 2015 if fossil energy use in a polluting sector were curtailed by a certain percentage (from 2\% to 20\%).

\noindent\textbf{Figure 3} Marginal damages measured in dollars attributable to an additional ton of CO$_2$ emissions from (a) rural and residential coal use, (b) coal use in the industry sector, (c) oil use in the industry sector, (d) coal use in the service sector, and (e) oil use in the transportation sector.

\noindent\textbf{Figure 4} Total damages from fossil fuel use in China (a) by sector and (b) by region with sectoral breakouts.

\noindent\textbf{Figure 5} Total damage by distance (measured by the edge length of the square centered at a monitor station) and emissions source type.

%\noindent\textbf{Figure 6} Sensitivity of population weighted PM$_{2.5}$ concentrations relative to fossil fuel use abatement in China.

\newpage
\clearpage
\section{Figures}

%\begin{figure*}[ht!]
%\centerline{\includegraphics[width=0.6\textwidth]{fig6.png}}
%\caption{Sensitivity of population weighted PM$_{2.5}$ concentrations relative to fossil fuel use abatement in China.}
%\label{fig:fig6}
%\end{figure*}

%TC:endignore

\end{document}

% ======================================================================

\begin{titlepage}
{\noindent\LARGE\bf\thetitle}

\bigskip

% : Insert author names, affiliations and corresponding author email
%FIXME
\begin{flushleft}\large
	Da Zhang\textsuperscript{1,2},
	Qingyi Wang\textsuperscript{3},
	Shaojie Song\textsuperscript{4},
	Simiao Chen\textsuperscript{5,6},
    Mingwei Li\textsuperscript{7},
    Lu Shen\textsuperscript{4},
	Siqi Zheng\textsuperscript{8},
    Bofeng Cai\textsuperscript{9,{*}},
    Shenhao Wang\textsuperscript{8,10,{*}}
\end{flushleft}

\bigskip

\noindent
%FIXME
\begin{enumerate}[label=\textbf{\arabic*}]
\item Institute of Energy, Economy, and Environment, Tsinghua University, Beijing, China
\item Joint Program on the Science and Policy of Global Change, Massachusetts Institute of Technology, Cambridge, MA, USA
\item Department of Civil and Environmental Engineering, Massachusetts Institute of Technology, Cambridge, MA, USA
\item School of Engineering and Applied Sciences, Harvard University, Cambridge, MA, USA
\item Heidelberg Institute of Global Health, Faculty of Medicine and University Hospital, Heidelberg University, Heidelberg, Germany
\item Chinese Academy of Medical Sciences and Peking Union Medical College, Beijing, China
\item Center for Policy Research on Energy and the Environment, Princeton University, Princeton, NJ, USA
\item Department of Urban Studies and Planning, Massachusetts Institute of Technology, Cambridge, MA, USA
\item Center for Climate Change and Environmental Policy, Chinese Academy for Environmental Planning, Beijing, China
\item Media Lab, Massachusetts Institute of Technology, Cambridge, MA, USA, 02139

%\item Department of Earth, Atmospheric, and Planetary Sciences, Massachusetts Institute of Technology, Cambridge, MA, USA, 02139
%\item Institute for Data, Systems, and Society, Massachusetts Institute of Technology, Cambridge, MA, USA, 02139
%\item Sloan School of Management, Massachusetts Institute of Technology, Cambridge, MA, USA, 02139
\end{enumerate}

\bigskip

% : Use the dagger symbol to denote a single equal contribution authorship.
% : Multiple equal-contribution authorship may be included in the acknowledgments.
%FIXME
%\textbf{{†}}: These authors contributed equally to this work.

% : Use the asterisk to denote corresponding authorship.
%FIXME
\textbf{*} To whom correspondence should be addressed. E-mail: shenhao@mit.edu and caibf@caep.org.cn.
\vfill

% ======================================================================
% : set word count results
% : to use `texcount` results, use '%TC:ignore'/'%TC:endignore' directives.
% : Exclude this part in submission
\wordcount{\wcAbstract}{\wcManuscript}{\wcMethods}

\end{titlepage}

\newpage

\section{Figure Captions}

\noindent\textbf{Supplementary Figure 1} Architecture of the ResCNN framework.

\noindent\textbf{Supplementary Figure 2} The distribution of marginal damages attributable to an additional ton of CO$_2$ emissions.

\noindent\textbf{Supplementary Figure 3} Prediction accuracy (measured by weighted $R^2$ on the train, validation, and test data sets) by models with different sizes of input area.

\noindent\textbf{Supplementary Figure 4} Prediction accuracy (measured by weighted $R^2$ on the train, validation, and test data sets) by model with output variables for different pollutant concentrations.

\section{Table Captions}

\noindent\textbf{Supplementary Table 1} Model prediction evaluation metrics (weighted by population of each observation) compared to existing studies.

\noindent\textbf{Supplementary Table 2} Seventeen dimensions of the hyper-parameter space.

\newpage
\begin{figure}[!h]
\begin{center}
\includegraphics[width=1.1\textwidth]{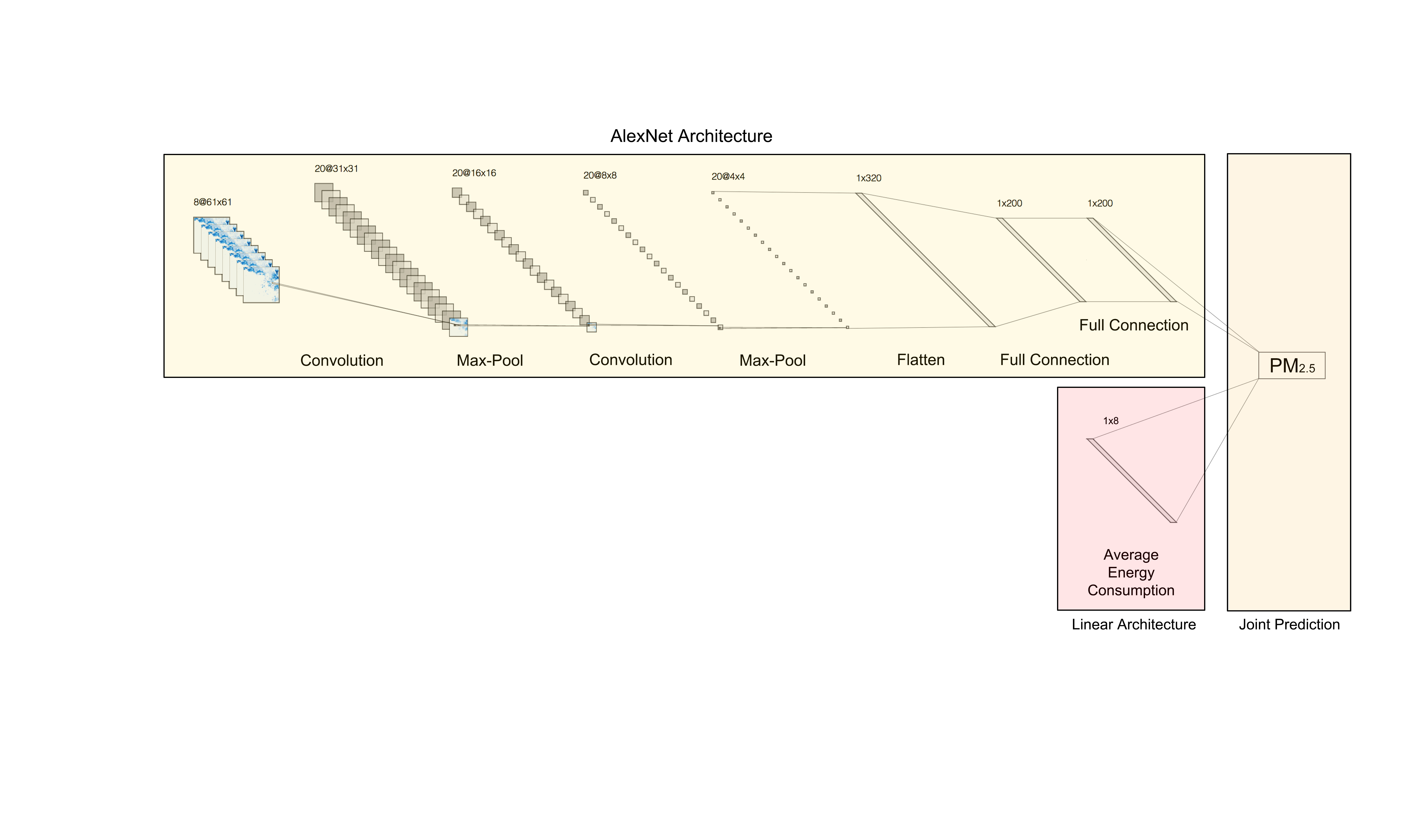}
\caption{Architecture of the ResCNN framework.}\label{fig:figS1}
\end{center}
\end{figure}

\begin{figure*}[!h]
\centering
\centerline{\includegraphics[width=1\textwidth]{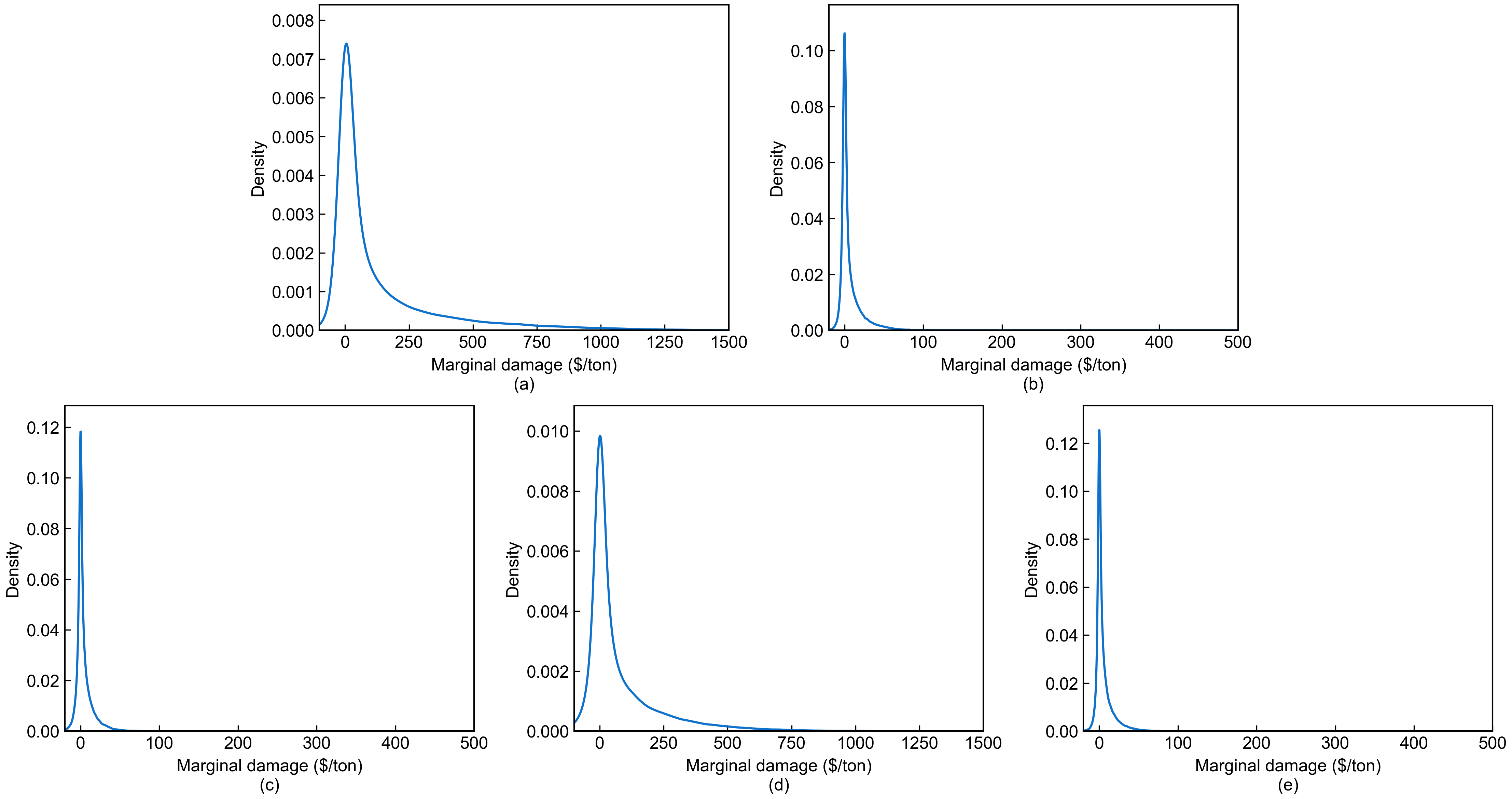}}
\caption{The distribution of marginal damages attributable to an additional ton of CO$_2$ emissions: (a) rural and residential coal use, (b) coal use in the industry sector, (c) oil use in the industry sector, (d) coal use in the service sector, and (e) oil use in the transportation sector.}
\label{fig:figS2}
\end{figure*}

\begin{figure*}[!h]
\centerline{\includegraphics[width=0.5\textwidth]{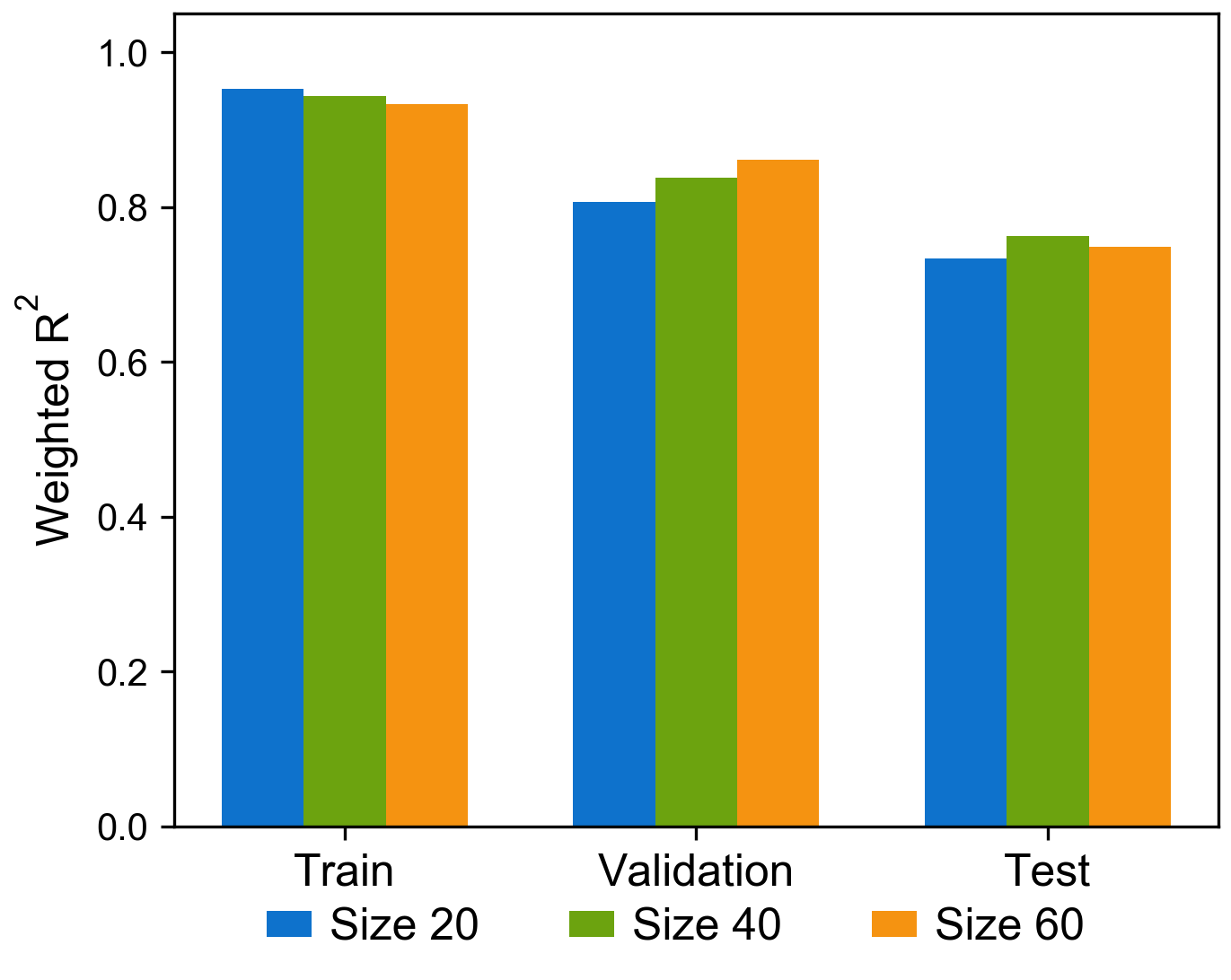}}
\caption{Prediction accuracy (measured by weighted $R^2$ on the train, validation, and test data sets) by models with different sizes of input area.}
\label{fig:figS3}
\end{figure*}

\begin{figure*}[!h]
\centerline{\includegraphics[width=0.6\textwidth]{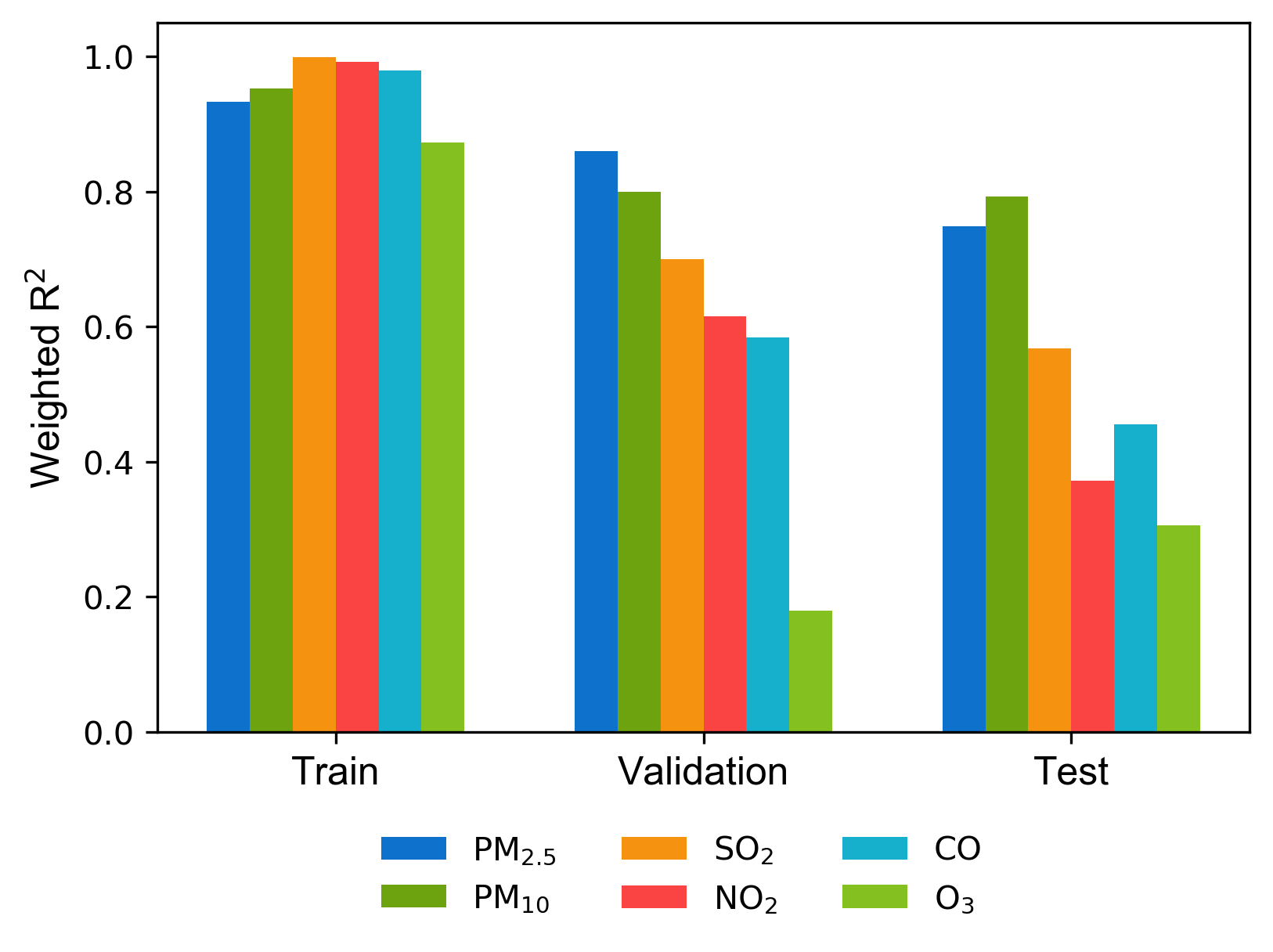}}
\caption{Prediction accuracy (measured by weighted $R^2$ on the train, validation, and test data sets) by model with output variables for different pollutant concentrations.}
\label{fig:figS4}
\end{figure*}

\newpage

\begin{table}[!h]
\centering
\caption{Model prediction evaluation metrics (weighted by population of each observation) compared to existing studies.}
\resizebox{1\linewidth}{!}{%
\begin{tabular}{p{0.35\linewidth} P{0.3\linewidth} P{0.2\linewidth}P{0.2\linewidth}P{0.2\linewidth}}
\hline
\textbf{Evaluation Metrics} & Equations & This study & InMAP & AP2 \\
 &  & (test data) &  &  \\
\hline
Mean fractional bias (MFB) & $\frac{1}{N}  \sum_n w_n \frac{2(y_n - \hat y_n)}{y_n + \hat y_n}$  & $-$0.04 & $-$0.06 & N/A\\
&  & & & \\
Mean fractional error (MFE) & $\frac{1}{N} {\sum_n} w_n \frac{ 2|y_n - \hat{y}_n|}{y_n + \hat y_n}$  & 0.12 & 0.36 & N/A\\
&  & & & \\
Mean proportional error (MPE) & $\frac{1}{N} {\sum_n} w_n \frac{|y_n - \hat{y}_n|}{\hat y_n}$ & 0.06 & N/A & 0.37 \\
&  &  &  & \\
Correlation coefficient (\rho) & $\frac{\sum_n w_n (y_n  \hat{y}_n) - \sum_n w_n y_n \sum_n w_n \hat{y}_n}{\sqrt{\sum_n w_n(y_n - \bar y_w)^2 } \sqrt{\sum_n w_n(\hat{y}_n - \bar{\hat{y}}_w)^2 }}$  &  0.88 & 0.74 & 0.62\\
&  & & & \\
$R^2$ & $\frac{{\sum_n} w_n (y_n - \bar{y}_w) (\hat{y}_n - \bar{\hat{y}}_w)}{\sqrt{\sum_n w_n(y_n - \bar y_w)^2 } \sqrt{\sum_n w_n(\hat{y}_n - \bar{\hat{y}}_w)^2 }}$  & 0.75 & 0.13 & N/A \\
%& \\
%Root Mean Squared Error (RMSE) & $\sqrt{\frac{1}{N} \sum_{n=1}^N w_n ( y_n - \hat y_n)^2 }$ \\
%& \\
%Normalized Root Mean Squared Error (NRMSE) & $\frac{\sqrt{\frac{1}{N} \sum_{n=1}^N w_n (y_n - \hat y_n)^2 }}{\bar{y}_w \times \bar{\hat{y}}_w} $ \\
%& \\
%Mean Bias (MB) & $\frac{1}{N} \underset{n}{\sum} w_n (y_n - \hat y_n)$ \\
%& \\
%Mean Error (ME) & $\frac{1}{N} \underset{n}{\sum} w_n |y_n - \hat y_n|$ \\
\hline
\multicolumn{5}{l}{Note: $y_n$ is the observed PM$_{2.5}$ annual average concentration, and $\hat{y_n}$ is the predicted PM$_{2.5}$}\\
\multicolumn{5}{l}{annual average concentration by a model. $w_n$ is the population weight. $\bar{y}_w$ is the weighted }\\
\multicolumn{5}{l}{ average of $y_n$, and $\bar{\hat{y}}_w$ is the weighted average of $\hat y_n$.}\\
\end{tabular}
} %resize used here
\label{table:metrics}
\end{table}

\begin{table}[!h]
\centering
\caption{Hyper-parameter Search Space.}
\resizebox{0.8\linewidth}{!}{%
\begin{tabular}{p{0.5\linewidth} p{0.3\linewidth} }
\hline
\textbf{Hyperparameters} & \textbf{Values} \\
\hline
Number of iterations & [100, 300, 500] \\
Size of mini batches & [20, 50, 100, 200] \\
Number of convolutional layers & [1,2,3,4,5] \\
Number of filters in convolutional layers & [20,50,80,100] \\
Kernel size in convolutional layers & [2,3,4,5,6] \\
Stride size in convolutional layers & [1,2] \\
Kernel size in max pooling layers & [2,3,4,5,6] \\
Stride size in max pooling layers & [1,2] \\
Existence of dropout layers & [True,False] \\
Dropout rates & [0.0,0.1,0.2,0.5] \\
Batch normalization & [True,False] \\ 
Number of fully connected layers & [1,2,3] \\
Width of fully connected layers & [50,100,200] \\
Existence of augmentation & [True,False] \\
$\epsilon$ variance in image augmentation & [0.05,0.1,0.2] \\
\hline
\end{tabular}
} %resize used here
\label{table:hyper_space}
\end{table}

\begin{table}[!h]
\centering
\caption{Pruned Hyper-parameter Search Space.}
\resizebox{0.8\linewidth}{!}{%
\begin{tabular}{p{0.5\linewidth} p{0.3\linewidth} }
\hline
\textbf{Hyperparameters} & \textbf{Values} \\
\hline
Number of iterations & [500] \\
Size of mini batches & [200] \\
Number of convolutional layers & [1,2] \\
Number of filters in convolutional layers & [20,50,80] \\
Kernel size in convolutional layers & [2,3] \\
Stride size in convolutional layers & [2] \\
Kernel size in max pooling layers & [2] \\
Stride size in max pooling layers & [2] \\
Existence of dropout layers & [True,False] \\
Dropout rates & [0.1] \\
Batch normalization & [True,False] \\ 
Number of fully connected layers & [1,2] \\
Width of fully connected layers & [200] \\
Existence of augmentation & [True,False] \\
$\epsilon$ variance in image augmentation & [0.05,0.1] \\
\hline
\end{tabular}
} %resize used here
\label{table:pruned_hyper_space}
\end{table}

% \clearpage
% \begin{thebibliography}{9}
% \bibitem{Burnett2014}
% Burnett R. T. et al.
% \newblock {An Integrated Risk Function for Estimating the Global Burden of Disease Attributable to Ambient Fine Particulate Matter Exposure.}
% \newblock {\em Environ. Health Perspect.} {\bf 122}, 397--403 (2014).
% \end{thebibliography}